\documentstyle[12pt]{article}
\title{On the Fractal Structure of the Universe}

\author{P V Gruji\'c$^\S$ and V D Pankovi\'c$^\dag$ \\
$^\S$ Institute of Physics, P.O. Box 57, 11000 Belgrade, Serbia\\
$^\dag$ Department of Physics, Faculty of Sciences, 21000 Novi
Sad,\\Trg Dositeja Obradovi\'ca 4, Serbia}

\date{}

\begin{document}
\baselineskip 15 pt

\maketitle \thispagestyle{empty}


\vskip.6cm

Short title: Fractal cosmos

\vskip.6cm

 \noindent
 {\bf  Abstract}. Despite the observational evidence that  the Universe  appears
 hierarchically structured up to a distance of at least 30 Mpc/h (and possibly up
 to 100 Mpc/h), the fractal paradigm has not yet been recognized by the majority
 of cosmologists today.  In this work we provide a brief overview of the recent
 observational and theoretical advances relevant to the question of the global
 cosmic structure and present some simple calculations which indicate how the
 hierarchical structure may pass over to the homogeneous Universe at very large
 scale. We show that the fractal structure may be derived from the moderately
 nonuniform matter distribution. We address a number of epistemological questions
 relevant to a general outlook of the Cosmos at large too.

\vskip2cm
         \vfill\eject\null

        \section{Introduction}
        \noindent

        Hierarchical models of the Universe appear as old as the
        cosmology. The first attempt to  describe an infinite cosmos
        was made by Anaxagoras, who conceived the world built
        according to the principle: {\it everything contains everything} (see,
         e.g. \cite{Grujic1}), what  is tantamount to assuming that each
         part of the physical world resembles the entire cosmos.
         This idea was taken up by a number of European cosmologists, from Kant
         to the present day ones (see, e.g. \cite{Grujic2}). The
         issue of the actual structure of our Cosmos appears
         controversial, the views going from one to the other extreme.
         The standard paradigm is that the Universe is spatially homogeneous
         and the inhomogeneities observed appear local  disturbances, which smear out
         on sufficiently large sale. The other extreme, at least formally, would
         be the assumption that the Universe has a hierarchical, fractal structure
         and the observational evidence of the large-scale homogeneity reflects our
         present day inability to  discern the higher-level
        hierarchical layers. Somewhere between these two end points
        lies the view that we have observational evidence about
        first two layers, galaxy clusters and superclusters, while the
        question concerning possible higher levels is to be left to
        the future observations.

              The possible hierarchical structuring of the Universe
              bears questions which go beyond the cosmological
              issue. The concept of fractality appears sufficiently general to
              be restricted to the mere arrangement of the galaxies in the observable
              Cosmos.  The very question of the cosmic structure brings in the old
              puzzle of the nature of the  space and time, which  intrigued
              cosmologists from the ancient time of Democritus, Aristotle to
              the modern naturalists, like Leibnitz, Berkeley, Mach, Einstein etc. In
              his recent work Roscoe argues that if one starts with the matter distribution
              in the Universe as the primary construct, prior to the determination of
              the space-time manyfold, one arrives at the fractal structure, with the
              fractal dimension $D_{f} = 2$  \cite{Roscoe}. The
              underlying principle of the fractal structure, the
              scale invariance, appears a powerful tool for examining fundamental
               properties on the microscopic level, like Schr{\"{o}}dinger equation,
               as shown by Brenig (\cite{Brenig}). Staying on the cosmic scale, Abdalla
               {\it et al} (\cite{Abdalla}) show that looking at the Cosmos within the
               past light cone, fractal structure appears even if the homogeneous
               distribution is assumed {\it a priory} (see, also \cite{Marcelo}).

                 The concept of fractality appears all but simple
                 one. On the phenomenological level,  observational
                 evidence of a possible hierarchical structuring at
                 large cosmic scale is difficult to confirm, for a
                 number of reasons. First, even if the overall
                 structuring is following a hierarchical pattern one
                 may hardly expect to  see an elegant, clear
                 geometrical picture.  The question of galaxy
                 distribution is essentially statistical one and one
                 needs to  do much elaborate study of the
                 astronomical catalogues in order to infer the
                 underlying regularities (see, e.g. \cite{Baryshev}).
                 Second, the galaxy distribution may follow a more
                 complicated pattern than simple fractal one, like
                 the multifractal ansatz, which may be described as
                 a non-uniform fractal distribution (see, e.g.
                 \cite{Galte}, but see, also \cite{Joseph}). Another
                 ingredient to be accounted for in studying the
                 large scale structure are cosmic voids, which  imply
                 their own scaling features (see, e.g. \cite{Jose},\cite{Yurij}).
                 Two main issues in studying the large scale cosmic
                 structures within the fractal paradigm are (i) what is the
                 exact value of the fractal dimension $\it D_f$ and
                 (ii) what is the value of the scale  factor $\it
                 \lambda_0$, separating the region with a clear
                 fractal distribution and the outer cosmic region, where
                 the galaxy distribution is definitely homogeneous
                 (see, e.g. \cite{Teerikorpi}, \cite{Yurij}).

\section{The fractal dimension}

   Accurate estimate of the cosmic fractal dimension appears of the
   crucial importance for several reasons. First, if different from
   $D_f = 3$ (fractal dimension must be less than the physical one, for
   topological reasons)the universe is endowed with a nontrivial structure,
   different from both uniform (homogeneous)and irregular inhomogeneity.
   Second, if the actual fractal dimension turns out to be an integer, it
   is possible to infer the geometry associated with  the dimension
   \cite{Teresa}. In particular, if it turns out that  $D_f = 2$ this would
   signal the presence of the holographic structure ("fractal holography")
   \cite{Mureika}. Recently, a particular interest in studying the actual metric
   of the observable universe by determining the matter distribution has
   arisen (see, e.g. \cite{Fulvio}, \cite{Mittal}).

     Hierarchical model allows for the cosmic accelerated expansion too,
     as demonstrated by a number of authors (see, e.g. \cite{Grujic3}),
     and the presence of dark matter (see, e.g. \cite{Francesco2}). Inclusion
     of the scaling symmetry, via renormalisation group,
     completes the collection of the symmetry principles, which serve
     as the most fundamental background of the cosmological models
     \cite{Grujic4}.

       Actual research, both observational and theoretical, revolves around
   the particular  $D_f = 2$ value, for several reasons, besides the
   possible holographic fractality. As shown by Charlier
    \cite{Charlier}, if Cosmos has $D_f \geq 2$ both Olbers'(blazing sky)  and
    Neumann-Seeliger's (gravitational) paradoxes are resolved
    \cite{Gabrielli}. Also, systems with such distribution have a
    compact projection onto a plane, like clouds shadows on Earth
    surface. In particular, hierarchical cosmic structure  provides
    an isotropic projection to the observers situated at an occupied
    point in the Universe, like our planet. That all observational
    evidences point towards $D_f = 2$ appears a significant
    empirical fact. This evidence, however, although does not rule
    out the hierarchical model, can not decide between two paradigms
     - homogeneous and hierarchical ones - either.

     Determining the fractal dimension and the scale where the fractal
     structure goes over to the homogeneous one appears all but easy task
     for the observational cosmology. Also, the usual procedures, based on the standard
     statistical method fail from the start, since they assume an approximately
     homogeneous distribution, which allows for making use of such quantities
     like the average density etc. By  such an approach hierarchical structure is
     ruled out from the beginning and the methodology requires a more general
     approach (see, e.g. \cite{Francesco}). Here we propose a simple
     model which allows the fractal structure to goes gradually to the
     homogeneous distribution. In the following chapter we present the
     model and in the last chapter a general discussion of the results and the
     prospects of the approach are given.

     \section{Calculations}

     \subsection{The model}

       Assuming  the galaxies equal mass points in the cosmic space,
       the most convenient way to infer the actual matter
       distribution from an occupied site (like our Milky Way) is to
       follow the mass (more precisely, the number of galaxies) as
       it increases with the radius R of a sphere centred on the observer.
       The formula used takes the form

        \begin{equation}
        N_r = a r^{D_f},  R_{min}\leq r \leq R_{max}, \label{eq: N_r}
        \end{equation}

       \noindent
       where $a$ is constant and $D_F$ is the fractal dimension. If
       $D_F = 3$ we have the usual formula for the uniform density.
       On the other hand, for $D_f < 3$ nonuniform distribution is
       present. As we mentioned before  $D_f = 2$ corresponds to a
       two-dimensional "space", as if the matter is uniformly
       distributed on the spherical surfaces. The question arises then
       as of the possible and/or actual values  the fractal dimension
       $ D_F$ assumes. One may put the question into a number of
       various forms.

       (i) Is there a unique value od $D_f$ for any possible scale
       (that is for $R_{max} \rightarrow \infty$)?
       (ii) If $R_{max}$ is finite, what is its value?

       Recent  observations have shown that for $R_{max}^{(1)} = 10 Mps/h$
       one has for the fractal dimension $D_f  \simeq 1.2$ \cite{Davies},
       \cite{Coleman}, whereas for $R_{max}^{(2)} = 100 Mps/h$ we saw that $D_f
       \approx 2$. These data indicate that the fractal dimension may be an
       increasing function of the scale, which eventually reaches
       its maximum value $D_f = 3$. We therefore  consider it
       appropriate to construct an analytical expression for the
       fractal dimension, which reproduces approximately the
       observed data.

\begin {equation}
   N = b \biggl(1 + \frac {R_{1}}{r} + \frac {R^{2}_{2}}{r^{2}}\biggr) r^3, \label{eq: N}
\end {equation}
\noindent where $N$ is the number of galaxies within a sphere of
radius $r$, $b$ is a constant, and $R_1 = R_{max}^{(1)}$ and $R_2
= R_{max}^{(2)}$ are bordering values between scales with
different fractal dimensions.

  We consider now three scale regions separately.

 (i) Innermost region:

\begin {equation}
 {R_2\over r} \leq ({R_1\over r})^2 = ({R_2\over 10r})^2, \label{eq: n1}
\end {equation},

 From (\ref{eq: N}) we have

\begin {equation}
   N = b R_1^3 r, \,\ r \leq {R_2\over 100}= {R_1\over 10},\label{eq: N1}
\end {equation}
\noindent
what provides $D_f = 1$.

 \vspace {.5cm}
(ii) Intermedium region
\begin {equation}
     R_2 \geq r \geq R_1 = {R_2\over 10}, \label{eq: n2}
\end{equation}
 From (\ref{eq: N}) we obtain

\begin {equation}
   N = b R_2^3 r^2, \,\ r \leq {R_2\over 100}= {R_1\over 10},\label{eq: N2}
\end {equation}

\noindent with $D_f = 2$.

 (iii) Finally, in the outmost space

\begin {equation}
   r \gg R_2 > R_1, \label{eq: n3}
\end {equation}

we have according to (\ref{eq: N})

\begin {equation}
   N = b r^3, \label{eq: N3}
\end {equation}
\noindent

and the "space" becomes finally Euclidian, with $D_f = 3$.

\subsection{The origin of fractal structure}

Much attention has been devoted to conceiving a mechanism
responsible for forming the hierarchical structure of the
observable universe (see, e.g. \cite{Grujic1}). Generally, one
may envisage two principal ways of forming fractal structure: (i)
breaking apart a primordial cosmic objects (so-called structuring
from above), and (ii) lumping together initial smaller units into
ever higher order systems (forming from below). Numerical
research of forming the fractal structure from a homogeneous
self-gravitating cosmic matter have been carried out (see, e.g.
\cite{Combes}). The common rationale for forming (any) structure
with galaxies as units is the presence of the Newtonian force of
the universal attraction.

   Within the model of an expanding Universe (Hubble flow), two
   competing effects are operative. Hubble flow tends to preserve
   (assumed)  initial, presumably homogeneous distribution, whereas
   mutual gravitational forces  force the galaxies two come together
    and form clusters etc. The later may join each other to form superclusters,
     and the process may continue (in principle) indefinitely (the so-called
     bottom-up mechanism, see, eg.g. \cite{Yoshida}. Of course,
     with the assumption that the age of the Universe is finite, there is no
     (cosmic)  time for forming too many hierarchical levels, and the net result
      of clustering would be the present existence of finite number of
      hierarchical levels, as observed.

   Generally, therefore, we take Newton gravitation as an destabilizing factor,
    whereas Hubble flow reduce the inevitable instability of the system with
    attractive forces only.

 We examine here possibility of deriving the hierarchical structure within
 the Newtonian gravitational dynamics applied to the approximately homogeneous
 statistical ensemble of galaxies.

We start with nearly homogeneously distributed statistical ensemble
of the galaxies with constant total mass $m$ and mass density
distribution $\rho$ within a sphere

\begin{equation}
\rho^{H}_{m}= \frac {3m}{4 \pi r^{3}}, \label{eq: rho}
\end{equation}

In the lowest, zero approximation order, with a completely
continuous and homogeneous distribution of $m/V$, absolute value of
the classical Newtonian gravitational force at the unit mass on the
sphere surface equals

\begin {equation}
    F = \frac {GM}{r^2} = \frac {G \rho^H_MV}{r^2}, \label{eq: F}                              .
\end {equation}

Now, we  take higher order approximation, which assumes discrete and
not quite homogeneous distribution of $m$ over $V$. It, formally,
can be done by the second order Taylor expansion (with absolute
value of any term and finite value of the variation $\delta r$ in
(\ref{eq: F}) of $\rho^{H}_{m}/r^{2}$ over $r$, with $V$ fixed. We
have

\begin {equation}
    \langle F\rangle = G \rho^{H}_{m}\frac {V}{r^{2}}+
    G \rho^{H}_{m}V |\frac {d(\frac {1}{r^{2}})}{dr}| \delta r +
    {1\over 2}  G \rho^{H}_{m}V|\frac {d^{2}(\frac {1}{r^{2}})}{dr^{2}}| \delta
    r^{2}, \label{eq: F1}
\end{equation}

\begin{equation}
  \langle F\rangle = G \rho^{H}_{m}\frac {V}{r^{2}}+ 2G \rho^{H}_{m}\delta r \frac
    {V}{r^{3}}+ 3G \rho^{H}_{m}\frac {V}{r^{4}}, \label{eq: F2}
    \end{equation}

    \begin{equation}
     \langle F\rangle = G \rho^{H}_{m}[1+ 2 \frac {\delta r}{r}  + 3(
    \frac {\delta r}{r})^{2}] \frac {V}{r^{2}}, \label{eq: F3}.
\end{equation}

From \ref{eq: F3} follows
\begin {equation}
   \rho_{m}= \rho^{H}_{m}[1+ 2 \frac {\delta r}{r} + 3(\frac {\delta r }{r})^{2}]
   \end {equation}

\noindent with the form compatible with (\ref{eq: N}).

\section{Conclusion}

Unlike the standard cosmological paradigm, which asserts that the
universe is homogeneous and isotropic, fractal paradigm is based
on  the so-called Conditional Cosmological Principle, as
formulated by Mandelbrot, which asserts that Cosmos appears
isotropic from any occupied point of space. This principle
appears less restrictive and more in  accordance with general
epistemology of science, which is particularly evident in the
case of  Quantum Mechanics and definitely less metaphysical. The
principle treats the observer and the object on equal footing,
what appears more acceptable than the usual "God's eye"
perspective.

 We have shown that partial fractal symmetry characteristic for (\ref{eq: N}) can be
simply obtained by classical, Newtonian gravitational dynamics,
applied to the nearly homogeneous statistical ensemble of the
galaxies. It practically means that partial fractal distribution
appears a part of the standard cosmological model
\cite{Teerikorpi}.

We have defined the partial fractal symmetry and have demonstrated
that it is neither totally global fractal (usual fractal) nor
totally local fractal (multi-fractal) symmetry. But, for discretely
different domains of the cosmic distances this partial fractal
symmetry can be effectively approximated by different global fractal
symmetric functions with corresponding discretely different fractal
dimensions. In this way, on the one hand, partial fractal symmetry
appears in a satisfactory agreement with astronomic data which
include practically all cosmic scales. We have shown that given
partial fractal symmetry can be obtained by classical, Newtonian
gravitational dynamics applied on the nearly homogeneous statistical
ensemble of the galaxies. In this sense, we show that partial
fractal cosmology appears a part of the standard cosmological model.

 \eject

\vfill
\end{document}